# Metal nanoparticle film-based room temperature Coulomb transistor

Svenja Willing[1,2,+], Hauke Lehmann[1,+], Mirjam Volkmann[1], Christian Klinke[1,*]

[1]*Institute of Physical Chemistry, University of Hamburg, 20146 Hamburg, Germany*

[2]*Deutsches Elektronen Synchrotron DESY, 22607 Hamburg, Germany*

[+] These authors contributed equally to this work.

* Correspondence should be addressed to klinke@chemie.uni-hamburg.de

**Abstract**

Single-electron transistors would represent an approach for less power consuming microelectronic devices if room-temperature operation and industry-compatible fabrication were possible. We present a concept based on stripes of small, self-assembled, colloidal, metal nanoparticles on a back-gate device architecture which leads to well-defined and well-controllable transistor characteristics. This Coulomb transistor has three main advantages: By employing the scalable Langmuir-Blodgett method we combine high-quality chemically synthesized metal nanoparticles with standard lithography techniques. The resulting transistors show on/off ratios above 90 %, reliable and sinusoidal Coulomb oscillations and room-temperature operation. Furthermore, this concept allows for versatile tuning of the device properties like Coulomb-energy gap, threshold voltage, as well as period, position and strength of the oscillations.

**One sentence summary**

A new transistor concept exploits the colloidal synthesis of metal nanoparticles and their Coulomb charging energy.

**Introduction**

In 1958, an article (*1*) reported on the unexpected observation of a negative differential resistance in Germanium p-n junctions. What started out as an anomaly in the known current-voltage characteristics of a diode led to the development of a new technology and resulted in a Nobel Prize for the author, Leo Esaki. Even beyond the technological point of view, studying tunnel processes enables the attainment of fundamental physical knowledge as Esaki mentioned in his Nobel Prize lecture (*2*). Today, especially single-electron applications (*3,4*) are a proof of how quantum mechanical tunnelling can enable new functionalities in conventional devices (*5*). Despite being similar to a common field-effect transistor, the single-electron transistor (SET) does not rely on the semiconductor band gap, but instead on the Coulomb-energy gap. The resulting transfer characteristics exhibit periodic on- and off-states known as Coulomb oscillations which might render new applications possible in the future. The working principle of the SET is well known both experimentally (*6-11*) and theoretically (*12*), and has been implemented in silicon devices (*13-16*). Nevertheless, employing this effect in an actual device that yields not only reproducible and predictable results, but can also be fabricated in an inexpensive, easy and industrially applicable way has remained a challenge. Wet-chemical synthesis is the best suited approach for the preparation of monodisperse nanoparticles with tuneable properties like size and inter-particle distance. They are solution-processable, stable and unstrained (*17,18*). However, reliably contacting single or multiple particles is not easily achieved. For example, bifunctional linker molecules (*19,20*), electrostatic trapping (*21*) and induced dipoles (*22*) are known approaches to catch a single nanoparticle or even assemblies thereof between readily fabricated electrodes and to demonstrate SET function (*23,24*). In semiconductor-nanoparticle arrays the coherence of the super-lattice determines whether

Coulomb charging or band-like transport governs the electrical transport (*25*). In extended metallic nanoparticle films the presence of the effect of Coulomb blockade can be extracted by analysis of the transport characteristics (*26,27*). In order to include high-quality nanoparticles as a 2D array into an electrical device, we use a unique combination of both top-down and bottom-up techniques. By means of the Langmuir-Blodgett method (*28-33*) the particles' tendency to self-assemble is exploited to create an ordered, homogeneous monolayer of evenly spaced nanoparticles. Position, shape and size of the monolayer can be easily and precisely controlled by a resist mask. With the Langmuir-Blodgett method as link between bottom-up and top-down, we can take advantage of the reliability and accessibility of physical techniques like lithography and sputter deposition for the general device fabrication while still profiting from the chemically synthesized and uniformly assembled high quality nanoparticles. Employing confined metal nanoparticle films as conductive channel in field-effect transistor geometry allows inducing Coulomb oscillations via a gate electrode. This work demonstrates an alternative device concept that renders thorough and systematic investigations possible. Understanding parameter dependencies and influences allows tailoring the devices to the needs of future applications and improvement in performance. We present Coulomb transistors with on/off ratios over 90 %, highly tuneable properties and an unprecedented clarity of Coulomb oscillations up to room temperature.

**Results and discussion**

*Device preparation*

Colloidal cobalt-platinum (CoPt) nanoparticles (Fig. S1) with diameters of $(2.3 \pm 0.2)$ nm, $(3.5 \pm 0.3)$ nm and $(4.0 \pm 0.3)$ nm were synthesized via the hot-injection method. CoPt was chosen due to the highly tuneable and well established synthesis (*34*) and because the high amount of platinum in the particles avoids oxidation and electromigration. The particle size can easily be adapted by varying the ratio of the platinum source and the amine that is used as one of two ligands. These ligands keep the particles solution-processable and their chainlength defines the particle spacing and, thus, the tunnel-barrier width. In our case, oleylamine and oleic acid were mainly used as ligands resulting in a particle spacing of about 1.9 nm (medium and small particles). Good results have also been obtained using the shorter chained versions decylamine and decanoic acid that lead to a separation of 1.4 nm (large particles). The electrodes were prepared on $Si/SiO_2$ substrates by a combination of electron-beam lithography including a marker-alignment procedure, metal evaporation of titanium and gold and sputter deposition of additional $SiO_2$ as gate dielectric (refer also to the Materials and Methods section at the end of the main text). The nanoparticle array was prepared and included onto the sample geometry by the Langmuir-Blodgett method (*28-33*) (setup in Fig. S2). For this, the particles are casted onto a liquid subphase and carefully forced together by barriers until they constitute a densely packed, hexagonally ordered monolayer (*30,35*) which is monitored via in-situ surface-pressure measurements. Lifting the former submerged substrate and thereby collecting the particles on the surface transfers the monolayer onto the devices. The electrical characterization was performed under vacuum atmosphere in a closed-cycle cryogenic probe station at temperatures between 4.5 K and room temperature. An exemplary nanoparticle monolayer as well as micrographs of the final devices and a scheme can be found in Fig. 1.

This fabrication route is not only simple and scalable but especially highly tuneable. On the one hand, the particles can be adjusted in size and shape as well as their inter-particle spacing (*17,18,30,32*). The monolayer, on the other hand, can be patterned by lithography to define stripe-shaped nanoparticle arrays. Adjusting length and width controls the total current as expected according to the varying number and length of possible conduction paths. Employing this stripe-shaped monolayer channel allows having full electrostatic control over all particles at once from the perpendicular direction without screening or stray currents (*33*). However, beyond that, the new device concept exposes further dependencies that can be used to adjust also working bias or periodicity: A local back-gate geometry is employed which allows fabrication of all electrodes prior to particle deposition. Consequently, the pristine ordering in the monolayer is preserved and interferences due to further preparation steps as well as trap states in a capping dielectric can be avoided (Fig. S3 - S4). Owing to the increased quality of transfer characteristics it was possible to observe clear Coulomb oscillations at about 210 K with the very same particles as in Ref. 33 where 150 K was the highest temperature. Furthermore, this enables a systematic analysis of the parameters' influence on the device performance. Amongst others, the effect of temperature, lateral bias voltage, dielectric thickness and gate position on the transport and on the device performance have been investigated.

*Coulomb blockade and electrical transport*

Applying a bias voltage to the nanoparticle channel shows that conduction in this system is not purely metallic but governed by tunnel barriers between individual particles. The transport is hindered due to the potential barrier given by the charging energy (*12*)

$$(1) \quad E_C = \frac{e^2}{2C}$$

required to overcome the electrostatic repulsion when an additional electron transfers onto a particle. The high charging energy results from the particles' inherent small self-capacitance $C$ which is proportional to the particle radius *r*. A detailed discussion of the capacitance as well as an estimation of its different contributions can be found in the Supplementary text S1. From this, a charging energy of about $E_C = 100$ meV is calculated for the medium sized particles. During transport, this Coulomb blockade leads to the emergence of discrete energy levels separated by a Coulomb-energy gap of $2 \cdot E_C$. Charge carriers can tunnel through the barriers to percolate through the monolayered channel. However, elastic tunnelling is highly sensitive to the energy difference between initial and final state (*36*) and, thus, to the electrons' energy given by the bias voltage and temperature. Synthesizing the nanoparticles by colloidal chemistry allows for a precise tailoring of the system's energy levels. Very small and uniform capacitances can be achieved with monodisperse nanoparticles. The tunnel-barrier width can be adjusted through the inter-particle spacing defined by the organic ligands.

Thorough investigations have been conducted with the medium sized particles of 3.5 nm in diameter and will be presented in the following if not stated otherwise. Fig. 2A and B display the output characteristics of an exemplary device. For low temperatures, compare the lower inset in Fig. 2A, no transport takes place up to a certain voltage that delivers the energy needed to overcome the Coulomb blockade. This so-called threshold voltage decreases with rising temperature, which is very well visible in the logarithmic plot of the respective differential conductances in Fig. 2B. Around 90 K (bold green graph), there is no observable Coulomb-blockade regime anymore as can be seen from the appearance of a finite zero-bias conductance. Even at room temperature the characteristic is still slightly non-linear. The monolayer system is well suited to investigate the temperature transition of the transport mechanism (*31-33*). In accordance with previous observations, electrical transport at low temperature takes place mainly through tunnelling (direct or Nordheim-Fowler tunnelling) where the current is strongly dependent on the source-drain voltage (*37*). At elevated temperatures, transport is increasingly due to thermally activated processes like electron hopping. The current-voltage dependencies of both mechanisms are exemplarily fitted to the data in the two insets in Fig. 2A (more detailed in Fig. S5).

In addition, the analysis of the temperature dependency of the differential conductance $G$ can give further information. The zero-bias conductance at temperatures T above the Coulomb blockade regime follows the equation

$$(2) \qquad G \propto \exp\left(\frac{1}{T^\nu}\right).$$

The parameter $\nu$ can take different values depending on the underlying transport type. While $\nu = 1/3$ and $\nu = 1/2$ would suggest variable-range hopping (*38,39*), our data is best fitted with $\nu = 1$. This Arrhenius type behavior can be seen in Fig. 2C and indicates nearest-neighbor hopping. In the case of Arrhenius behavior, the differential zero-bias conductance

$$(3) \qquad G \propto \exp\left(-\frac{E_A}{k_B T}\right)$$

can be used to determine the system's activation energy $E_A$. For the medium sized particles (r = 1.75 nm) this yields $E_A = 61$ meV. The dependency on the particle size will be discussed later. Applying a lateral bias voltage will lower the effective barrier for the charge carriers and, thus, the activation energy. Using the differential conductance at non-zero source-drain voltage values and fitting the temperature dependency as before allows obtaining these activation energies. As shown in Fig. 2D, the activation energy decays roughly exponentially with the applied electric field independent of channel length or gating geometry. Overall, the transport investigations support the monodispersity and super-crystallinity of the particle monolayer.

*Coulomb Oscillations and Tunability*

An additional gate electrode separated from the channel by a dielectric layer allows influencing the transport capacitively. Tuning the gate voltage results in a field effect that continuously shifts the energy levels of the particles. Whenever neighboring energy states align, tunnelling is most probable and a maximum current flow can be observed. Exemplary transfer characteristics are displayed in Fig. 3 (temperature dependency) and Fig. 4 (source-drain voltage dependency, a 2D pseudo-color plot can additionally be found in Fig. S6 in the Supplementary Materials). They show these periodic Coulomb oscillations with, for extended particle systems, unprecedented regularity and clarity (32,33). To quantify the results, the relative on/off ratios between main minimum and neighboring maximum, for consistency always on the right side, have been calculated and can be found in the graphs as percentages.

Sweeping the potential of a local 0.5 µm wide back gate below a 1 µm long conduction channel resulted in the transfer characteristics shown in Fig. 3A. It can be seen, that the current clearly oscillates and that there is only little noise and hysteresis between the two sweep directions. All main features reproduce for the different temperatures and there is one pronounced minimum, which will be discussed in detail later. At low temperatures, on/off ratios of about 70 % could be observed with the medium sized particles. With increasing temperature, the electrons' energy and, thus, the overall current is higher as more and more conduction paths are accessible for transport. However, this multitude of conduction paths cannot be switched as efficiently simultaneously due to statistical variations of threshold voltage and Coulomb energy. This results in a quasi-constant background current and the relative on/off ratio decreases exponentially with temperature as displayed in Fig. 3B. Three devices are compared differing only in channel length. Although the differences are not large, there is a tendency that shorter channels yield higher on/off ratios with the same sized back gate.

Employing the global silicon substrate as back gate, which stretches over the whole channel length in contrast to the local gate, yields Coulomb oscillations with even higher regularity and periodicity (Fig. 3C). The transfer characteristics appear sinusoidal and there is no pronounced main minimum anymore. The device performance is about the same compared to the local gate case (Fig. 3A). It is worth noting, that clear oscillations can be observed at temperatures of up to at least 240 K. Beyond that, both gating geometries show evidence of room-temperature Coulomb oscillations. However, the obtained percentages were negligible.

To improve the device performance even further, the adjustability of the system can be utilized in yet another way. By decreasing the particle size, the self-capacitance is lowered and the charging energy is increased. Hence, Coulomb oscillations become more pronounced and remain visible up to higher temperatures. Fig. 3D shows results from a device with smaller nanoparticles of 2.3 nm in diameter gated again by the global substrate. At the lowest measured temperature of 6 K, a relative on/off ratio of over 90 % could be obtained. Moreover, sinusoidal oscillations were clearly detectable at room temperature. The on/off ratio was six times higher than observed before and values of up to 5 % could be measured.

Further investigations demonstrate the influence of the bias voltage on the oscillations at 10 K (Fig. 4A; the chosen voltage values are also marked by the vertical lines in the respective output characteristics in Fig. 4B, more data is given in Fig. S7). The lowest displayed bias voltage of $V_{SD}$ = -9 V is clearly still inside the blockade regime. Nevertheless, weak oscillations can be seen in the red curve. Although the signal is very noisy, this shows how the gate's field effect can lower the energy barrier, so that current flow is periodically possible. Increasing the bias voltage enables more conduction paths. However, similar to the effect of a high temperature, the paths all vary slightly in their energy level configuration and cannot be blocked simultaneously leading to lower relative on/off ratios. Thus, the best ratios were obtained for bias voltages near the threshold voltage as depicted in Fig. 4C. This subfigure summarizes exemplary device performances at a temperature of 6 K for different bias voltages. For each curve, there is a blockade regime, then an onset of oscillatory behavior, a maximum of the on/off ratio around the threshold voltage, followed by a slow decrease. The longer the conducting channel, the higher the respective oscillation onset and threshold voltage, because the total bias is distributed over more particles. Again it can be seen, that the best performance was observed in the device with the shortest channel.

The four oscillations shown in Fig. 4A are shifting towards more negative gate voltages as the applied bias changes in the same direction. In the pronounced main minimum of the gate-voltage sweep not only the energy-level configuration is most unfavorable for transport but also the overall electric-field gradient between the three contacts is disadvantageous for elastic tunnelling. Depending on the exact location of the gate electrode under the channel and on the gate width, a simple model (*32*) suggests that this main minimum is at $V_{gate} \approx V_{SD}/2$. At this gate voltage, the linear potential gradient is least disturbed and remains very unfavorable for resonant tunneling so that the current is minimal. The potential gradient between source and drain is simulated and evaluated for two different gate positions based on this model in Fig. S8. To investigate this effect in more detail, structures with three parallel narrow local back gates have been fabricated as can be seen in the optical micrograph in Figure 5A. They have the same width, are separated from the channel by the same amount of dielectric and, therefore, only differ in their lateral placement and influence the transport along the channel from different positions.

Applying the same bias voltage to source and drain and using each one of the gates alone while leaving the other two uncontacted (floating) results in a shift in the oscillations depending on the used gate's position of about 16 V (Fig. 5B). It is now possible to use the gate voltage corresponding to the oscillation minimum as marker for the shift and plot it for different bias voltages (Fig. 5C). Independent of the temperature, there is a linear correlation between the gate voltage at the minimum and the applied bias voltage. It was found that the slope of this line can directly be connected to the gate's position. The theoretical slopes from simple simulations (Fig. S8) are also given in Fig. 5C. Although they are all somewhat smaller than those obtained from the data, the difference is about the same for all three lines. This leads to the conclusion that there might have been a shift in the lithography process and the gates are positioned slightly more to the right than intended. Note, these gates are still quite wide in comparison to the channel. The correlation between gate placement and oscillation is valuable for applications: The position of a narrow local gate can directly be used to shift the resulting transfer characteristics, or *vice versa*, the shift can be used to determine the gate position. For an application that requires the working voltage to be at a certain point of the oscillation, e.g. in the minimum for current stabilization, this offers an easy accessible tuning mechanism to adapt the functionality even if the bias voltage is given. Extending the study with the three gates to different channel lengths (Fig. S9) confirms the tendency that a better effect is obtained if the gate stretches over a larger fraction of the channel and adds that a gate electrode has no influence anymore if it is placed right next to the channel underneath the gold leads. In summary, the above observations show that depending on the focus of the desired application, one could either employ a narrow local gate to tune the position of the oscillations, or use a wide gate that covers the whole channel area to maximize the on/off ratio.

The device geometry determines many of the system's characteristic properties, however, the nanoparticle diameter has an even stronger influence. The usage of wet-chemically synthesized, colloidal nanoparticles makes small sizes and precise tailoring possible. In Fig. 6A, the size distributions of three batches of particles are shown. They range between 2.3 nm and 4.0 nm: The smaller the particles, the higher the self-capacitance and, thus, the charging energy. This can be seen directly in the transfer characteristics: Sweeping the gate voltage from one maximum to the next means adding the required energy to overcome the corresponding Coulomb-energy gap. The period of oscillation becomes longer for smaller particles. This correlation can very well be seen in Fig. 6B. For each particle size, a point cloud of oscillation-period measurements is given. It was found that the period did not depend on channel length, temperature or bias voltage, which is in line with theory. The horizontal lines mark the average value also given as a number on the right. There is also a second influence on the periodicity: The thickness of the dielectric material, meaning the separation of channel and gate electrode, determines the strength of the electric field in the nanoparticles. A thinner dielectric strongly increases the field effect and also shortens the oscillation period (compare red lines in Fig. 6B). While a thin dielectric also improves device performance, it still needs a certain thickness to act as barrier against leakage currents.

Increasing the Coulomb energy by decreasing particle size also leads to more pronounced oscillations. They remain visible up to a higher temperature and exhibit higher on/off ratios. Fig. 6C summarizes device performance at the lowest measured temperature for the three particle sizes. Simply changing from the medium particles to the small ones increased the on/off ratio by about 15 % and enabled sinusoidal room-temperature oscillations. As discussed in the beginning of this work, the system's activation energy can be determined from Arrhenius plots of the differential zero-bias conductance. Being strongly related to the Coulomb energy itself, the activation energy is also proportional to the inverse particle diameter. In Fig. 6D, the three data points from particles investigated in this work extend the data from previous work (*31*). The linear fit matches the relation very well and confirms the expected behavior.

Fabricating the gate electrode prior to particle deposition leaves the particle array that constitutes the conducting channel undisturbed and the Coulomb oscillations can be observed with little interference. Therefore, the influence of different gating techniques and parameters could be reliably investigated. It has been observed, that even the sputter deposited silicon dioxide underneath the particles can influence the transfer characteristics. Charging of trap states in the gate dielectric leads to some reversible hysteresis effects if the applied gate voltage is very high (see Fig. S3 in the Supplementary Materials). For later applications, some form of sealing will be required. It was found, that polymer systems like a resist or wax do not interfere with the devices function (see Fig. S4).

**Conclusion**

We demonstrate a new device concept that allows for easy and industry-compatible fabrication as well as a high tunability to match the needs of transistors or future applications. The device parameters can influence especially working temperature and bias, as well as oscillation period and strength. The use of colloidal nanoparticles ensures the tunability of size and tunnel-barrier width, the monodispersity, the very small capacitances and thus high Coulomb energies, and the inexpensiveness. In the metal nanoparticle based devices Coulomb oscillations are observed with an unprecedented clarity and reliability, even at room temperature.

**Materials and Methods**

*Nanoparticle Synthesis*

CoPt nanoparticles were synthesized according to the work done by Shevchenko et al. (*40*) with some modifications. The mainly investigated particles are monodisperse with a diameter of $(3.5 \pm 0.3)$ nm and a composition of 93,2 % Pt and 6,8 % Co (analysis of the (111) XRD peak by means of Vegard's law). In a first step, 33 mg (0.084 mmol) of the platinum source, platinum(II) acetylacetonate ($Pt(acac)_2$), are dissolved in 4 mL of the solvent 1-octadecene. To assist in the reduction of the platinum, 131.9 mg (0.51 mmol) of hexadecandiol is added. For shape and size control during particle nucleation and growth 2.22 mL (6.72 mmol) of oleylamine and 0.44 mL (1.4 mmol) of oleic acid are added. The solution is heated to 80 °C and stirred at this temperature for at least an hour under vacuum conditions. In the second step, 43.7 mg (0.126 mmol) of the cobalt source, dicobaltoctocarbonyl ($Co_2(CO)_8$), dissolved in 0.6 ml of 1,2-dichlorobenzene in an ultrasound bath, is injected rapidly into the solution that has been heated to 160 °C under nitrogen atmosphere. This so-called hot injection is followed by thermal decomposition of the cobalt precursor. The mixture immediately turns black and is continuously stirred at the injection temperature for two hours. During cooling down 5 mL of toluene are added. Then, the particles are washed three times through precipitation with 2-propanol and methanol and phase separated in a centrifuge with 7000 rpm (4492 g) for 4 min and subsequently resuspended in toluene. In a final step, they are filtered through a PTFE syringe filter with a porousness of 0.2 μm. Other particle sizes can be realized by altering the ratio of amine and platinum source. Various inter-particle distances can be achieved by substituting longer or shorter chained ligands.

*Sample preparation*

Doped silicon wafers of roughly 1 cm$^2$ with 500 nm thermal oxide are used as substrates. For electrode preparation, they are covered with the positive resist poly(methyl-methacrylate) (PMMA, dissolved in chlorobenzol) via spin-coating. Standard electron-beam lithography (Quanta SEM from FEI and CAD software ELPHY Plus from Raith) and metal evaporation of 2 nm titanium as adhesion layer and 23 nm of gold create the electrodes and contact pads as well as markers to align the different layers during lithography. As a gate dielectric, an isolating layer of about 130 nm silicon dioxide is deposited via ion-beam sputtering (PECS Modell 682 from Gatan) on top of the back-gate electrode while the contact pad is protected with PMMA that can later be removed. In a last step, another resist mask is created that defines the conduction channel as stripe between source and drain electrode. Some samples did not have a local back gate, but were used by employing the global silicon substrate as gate. In these cases, the pristine thermal oxide was only 300 nm thick.

*Langmuir-Blodgett Method*

Monolayer preparation was done using the Langmuir-Blodgett trough from KSV NIMA seen in Fig. S2. The silicon substrates with readily fabricated electrodes are attached to a holder together with a TEM grid to investigate the film quality afterwards. They are submerged into the subphase diethylene glycol. The particles are washed again, dissolved in toluene and carefully brought upon the subphase surface with a microliter syringe (start with 300-400 μL of nanoparticles in toluene as described under synthesis for trough area of 120 cm$^2$). Due to the hydrophobic character of the ligand shell they will stay on the surface and spread as far as possible while the solvent evaporates. With two computer controlled barriers they are slowly forced together with a speed of 1 mm/min, the compression is monitored by in-situ surface-pressure measurement. The slope changes of the resulting isothermal curve as well as empirical known values of the required difference in surface pressure enable estimating the point of a densely packed monolayer without holes or overlap. The barriers keep the film at this pressure for 2 hours relaxation time. Then the substrates are carefully lifted and slowly break the surface at an angle of 105° that allows collecting the particle film (*30*).

*Particle and sample characterization*

The CoPt nanoparticles were investigated with a transmission electron microscope (TEM) after synthesis and after film preparation. During each Langmuir-Blodgett process, a TEM grid was loaded with particles in parallel to the actual silicon substrates. From these micrographs the particle diameter, size distribution, spacing and monolayer quality were assessed. Exemplary batches of particles were measured with X-ray diffraction to verify the crystal structure and composition. All throughout the fabrication process the substrates, electrodes, resist masks and the final sample are frequently checked with a light microscope. Furthermore, in exemplary samples, the heights of the electrodes as well as of the sputtered dielectric are measured with an atomic force microscope.

*Electrical characterization*

The main measurements are carried out in the closed-cycle Cryogenic Probe Station (Model CRX-4K from LakeShore Cryotronics) with a Keithley 4200-SCS parameter analyzer. Characterization is done at residual gas pressures of $10^{-6}$ mbar at room temperature to $10^{-7}$ mbar at low temperature. The sample stage can be cooled down to 4.5 K. Contacting is done with up to six micro-manipulatable needle probes with a 20 µm radius beryllium-copper tip. Another pin is connected to the sample stage and can be used to apply a voltage to the substrate and, thus, to the global back gate. Signal transmission is achieved through triaxial cables to the Keithley's source measure units. Currents down to about 100 fA can be detected, while everything below will be considered as noise. For output characteristics the gate is left floating or grounded while the bias voltage is varied stepwise between ±20 V. Similarly, for transfer characteristics the gate voltage is varied stepwise in different intervals depending on the sample with a fixed bias voltage. Parameter sweeps are always done bi-directional to verify the results.

**Acknowledgments**

The authors thank the European Research Council for funding an ERC Starting Grant (Project: 2D-SYNETRA (304980), Seventh Framework Program FP7). The German Research Foundation DFG is acknowledged for financial support and granting the project KL 1453/9-1 and the Cluster of Excellence "Center of ultrafast imaging CUI".

Almut Barck is acknowledged for XRD measurements.

The authors declare that they have no competing interests.

Author Contributions: S.W., H.L., and C.K. conceived the main concepts. S.W. and H.L. performed e-beam lithography and the electrical measurements. M.V. performed the nanoparticle synthesis and characterization. S.W. and M.V. prepared the Langmuir-Blodgett films. C.K. performed the simulations. S.W., H.L., M.V. and C.K. analyzed the measurement data, discussed the results and wrote the manuscript.

All data needed to evaluate the conclusions in the paper are present in the paper and/or the Supplementary Materials. Additional data available from authors upon request.

**List of Supplementary Materials**

Supplementary Text S1

Supplementary Figures S1 to S9

**Figures**

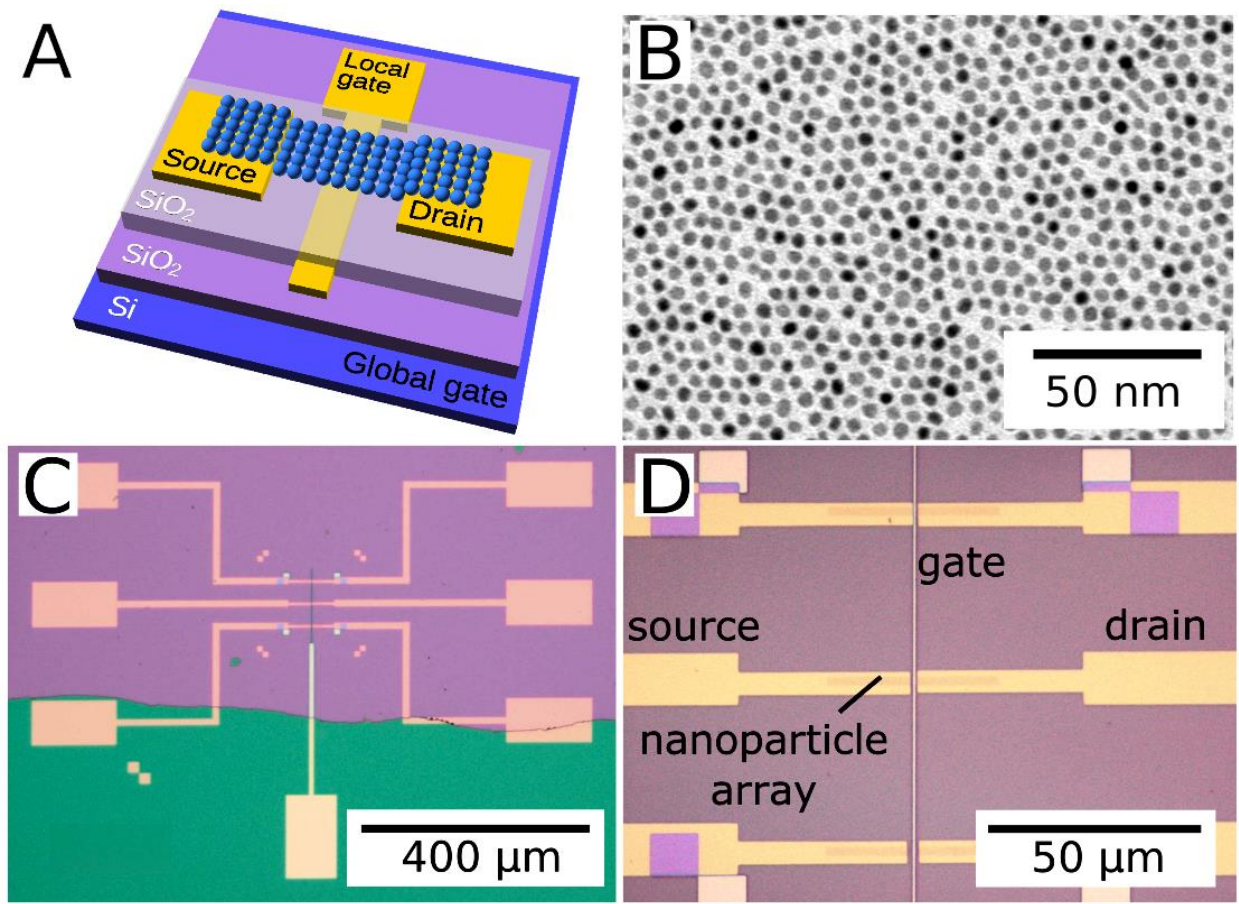

**Figure 1. Device design.** (**A**) Schematic drawing. (**B**) Transmission electron micrograph of the CoPt nanoparticle monolayer as assembled by the Langmuir-Blodgett method. (**C** and **D**) Light micrographs of final devices with three pairs of source and drain contacts as well as a local back-gate electrode. The gap between the leads and, thus, the length of the nanoparticle channel is 1, 2 or 3 µm (top to bottom).

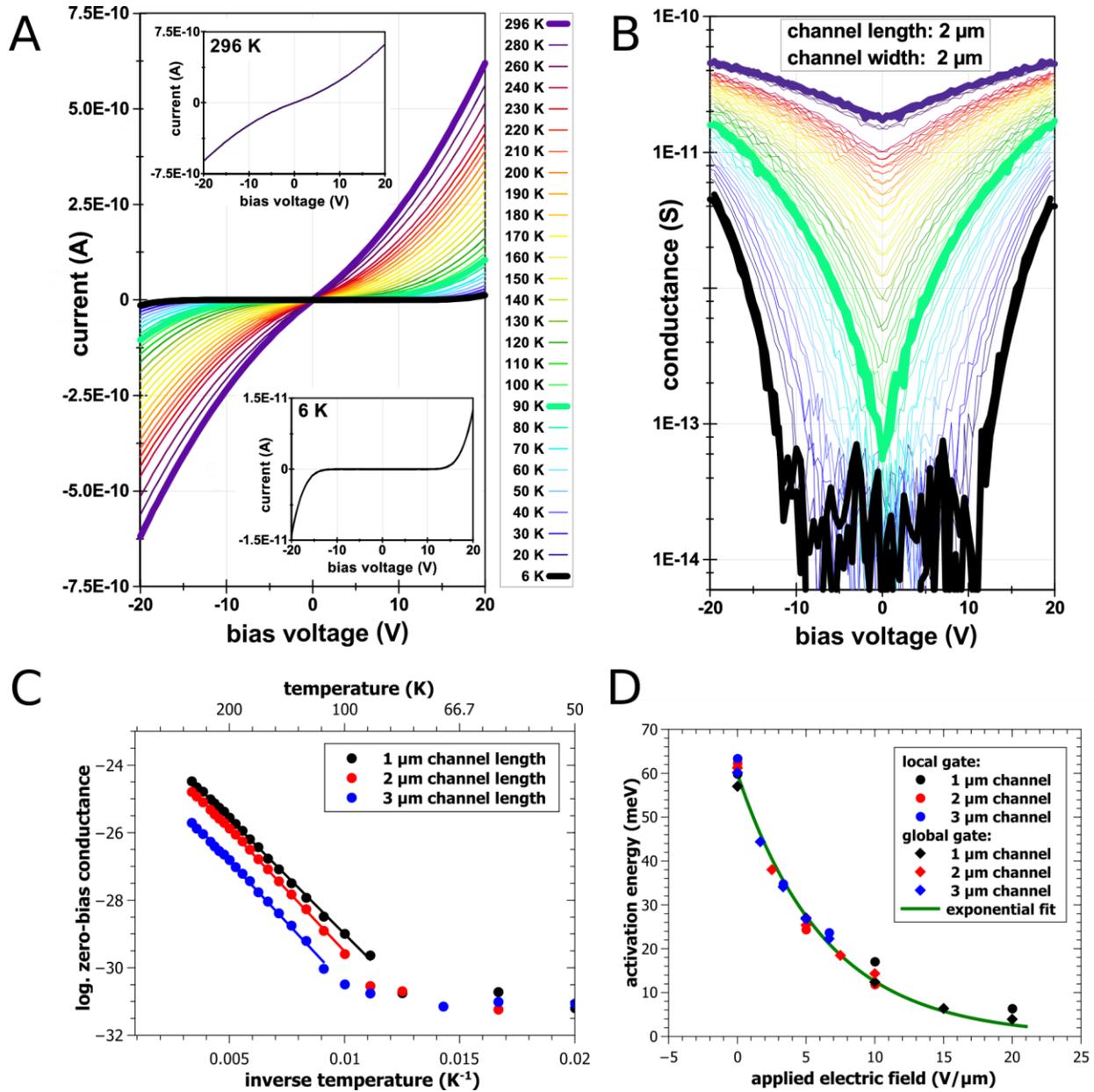

**Figure 2. Output characteristics.** (**A**) Current-voltage curves for different temperatures show the transition from a system governed by Coulomb blockade to almost Ohmic conduction. Lowest and highest temperatures are included as insets to provide a good comparability and are fitted with the respective transport mechanism. (**B**) Same data plotted logarithmically as differential conductance over bias voltage. The decrease of the blockade-regime width, the changes in threshold voltage and the increase in zero-bias conductance (at 90 K and above) can be seen. (**C**) The Arrhenius plot allows for a linear fit of the logarithmic zero-bias conductance outside of the Coulomb-blockade regime. The slope indicating the activation energy is similar for all channel lengths. (**D**) The activation energy decreases exponentially with the applied electric field and is not depending on channel length or gating geometry.

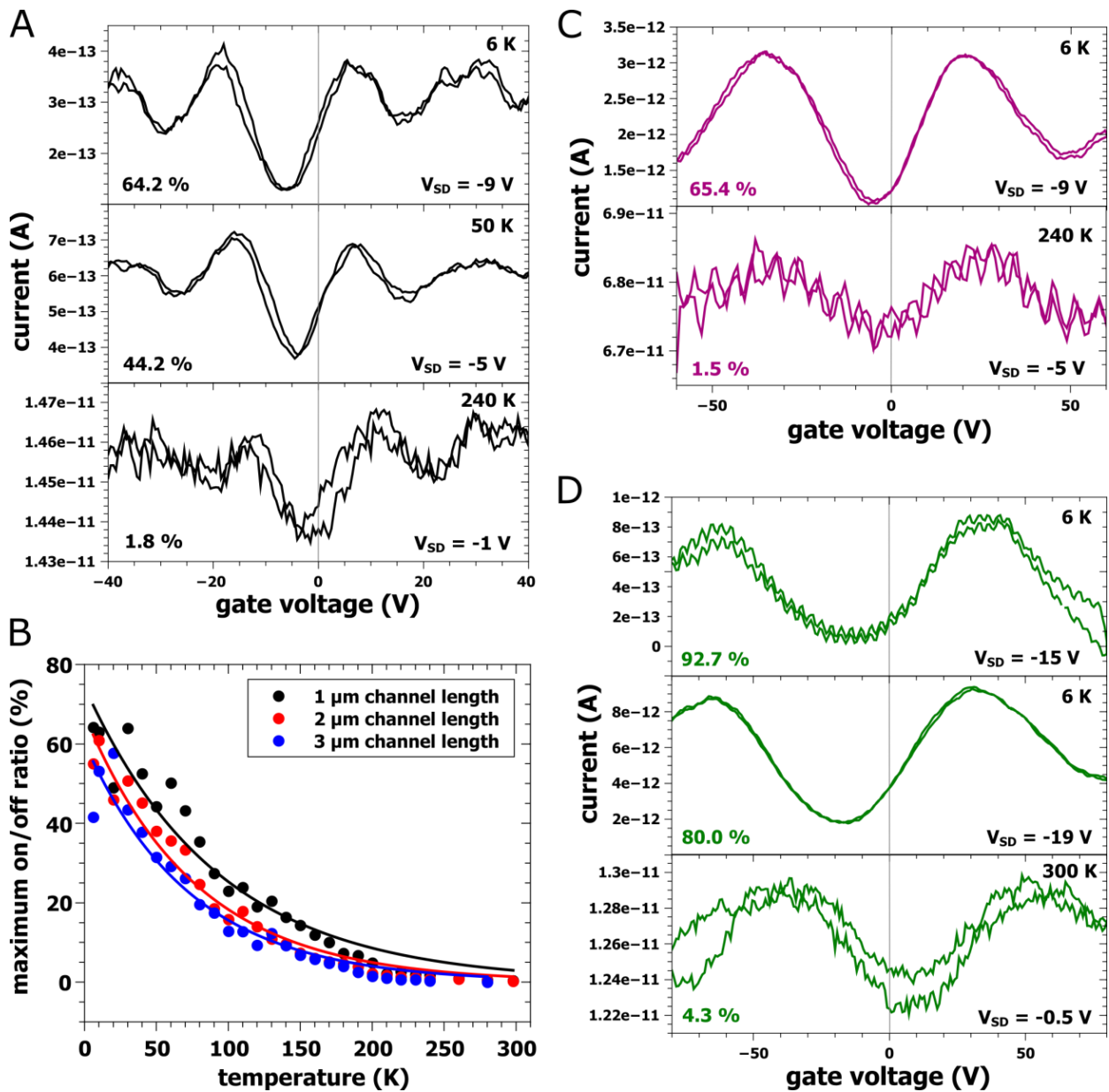

**Figure 3. Temperature dependency.** (**A**) The local back gate induces clear oscillations with reproducing features during a bi-directional gate-voltage sweep for different temperatures. (**B**) Highest found on/off ratios for three different channel lengths measured with a locally gated sample decay exponentially with increasing temperature. (**C**) A sample gated with the global substrate shows very periodic oscillations with similar percentages. (**D**) Smaller nanoparticles (2.3 nm in diameter) lead to higher on/off ratios and more pronounced oscillations at higher

temperatures. Minor changes in bias voltage adjust the current and render the curve perfectly smooth. Even above room temperature clear Coulomb oscillations are visible and efficiencies of 5 % could be reached. A larger distance to the gate electrode and smaller particles elongate the oscillation period from an average of 24.3 V (A) to 54.7 V (C) and 96.9 V (D) respectively. Note that the origin of the higher frequency oscillations seen in C and D is not entirely clear, they are assumed to result from some external influence and only appeared from time to time.

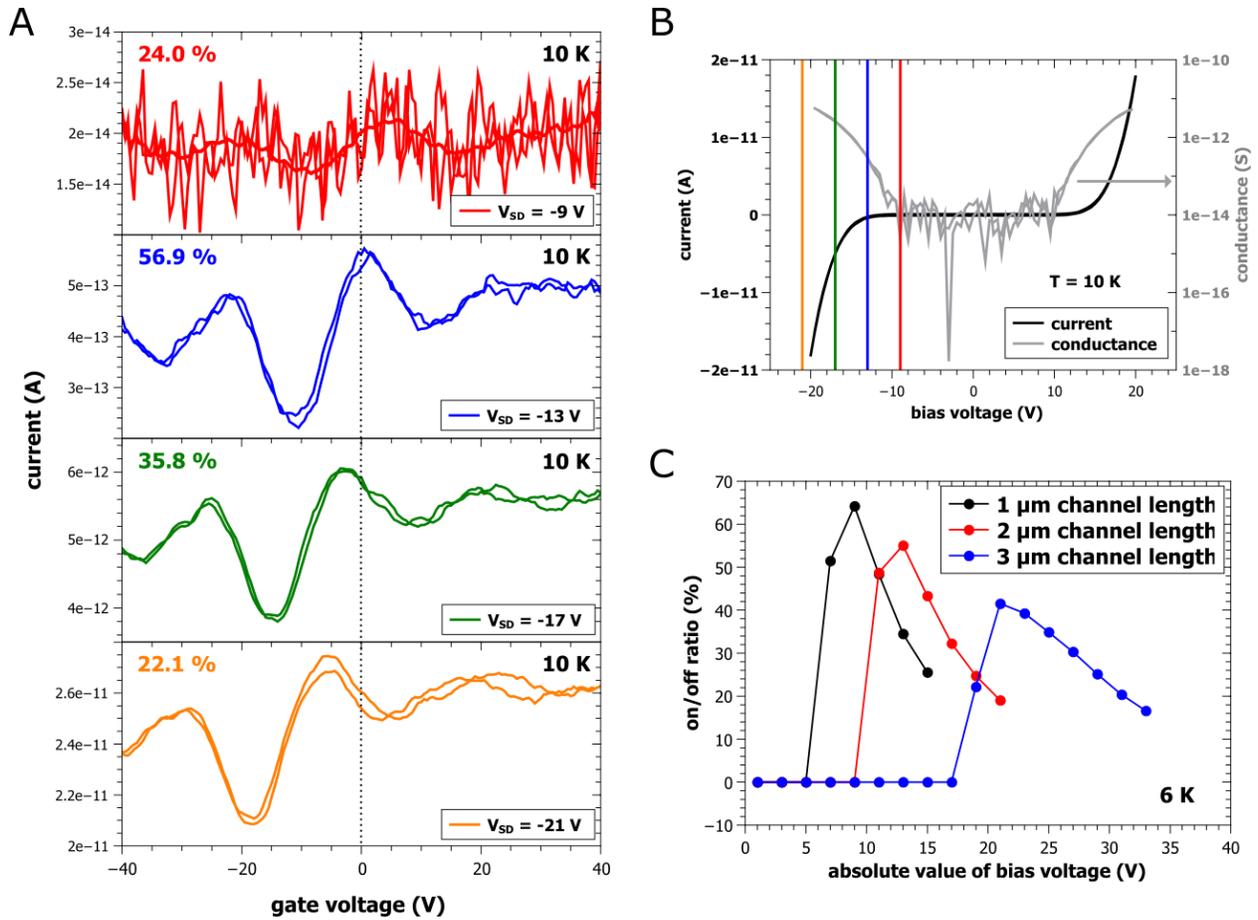

**Figure 4. Voltage dependency.** (**A**) Oscillations induced by a local gate for different bias voltages at a temperature of 10 K. For the lowest shown voltage, an averaged curve has been added as guide-to-the-eye. The shape and distinctive features of each curve are reproducing at different bias voltages. A shift in minimum position can very well be seen. (**B**) Corresponding output characteristics and differential conductance plot. The colored lines mark the bias voltages displayed in (A). (**C**) The bias-voltage dependency of the on/off ratio illustrates the different stages of electrical transport in the system. The maximum performance is reached around the threshold voltage.

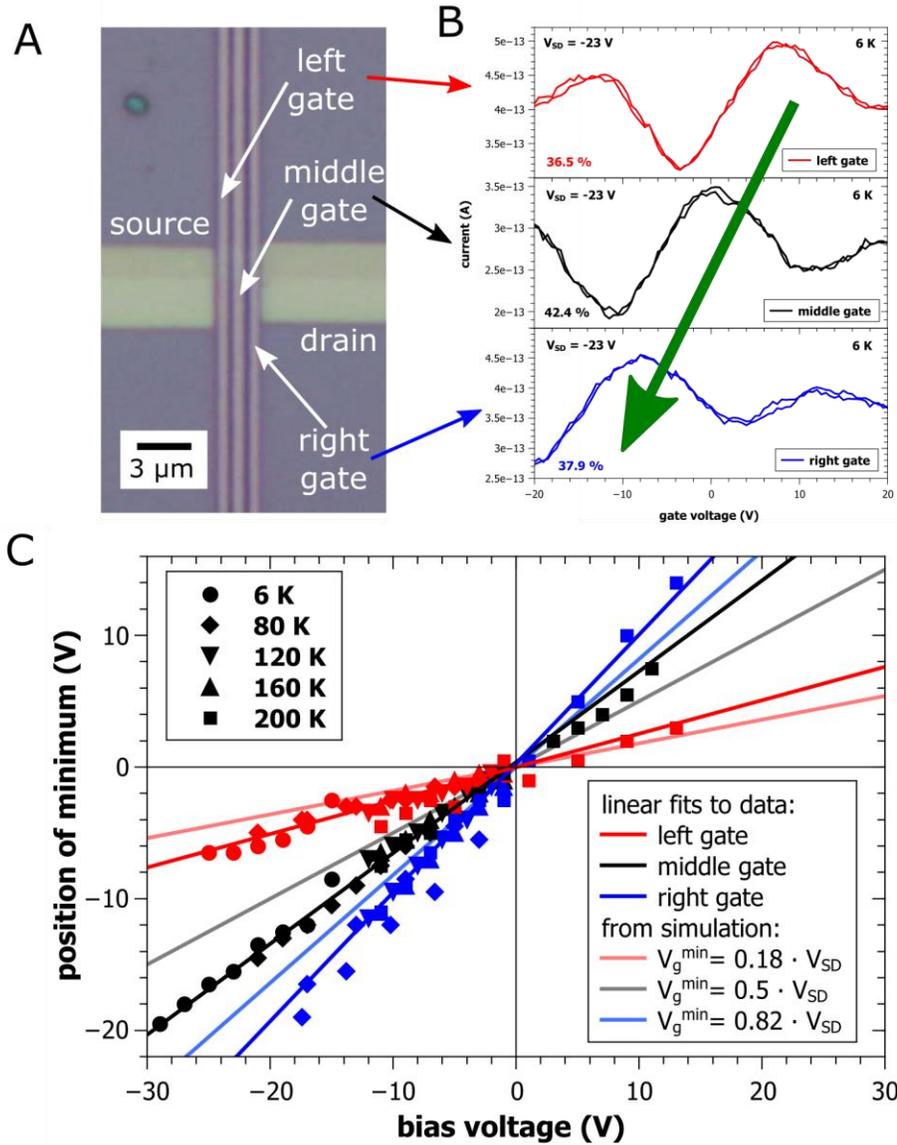

**Figure 5. Oscillation shift.** (**A**) Optical micrograph of a device with a 3 μm long channel and three parallel 500 nm wide gate electrodes. (**B**) Each gate can be used to induce oscillations. At the same bias voltage the main features are shifted. (**C**) Plotting the position of the minimum with respect to the applied bias voltage yields a linear correlation. It is independent of temperature. Each gate position corresponds to a certain slope. The measured data roughly matches the simulated slopes but reveals a shift in the lithographic alignment.

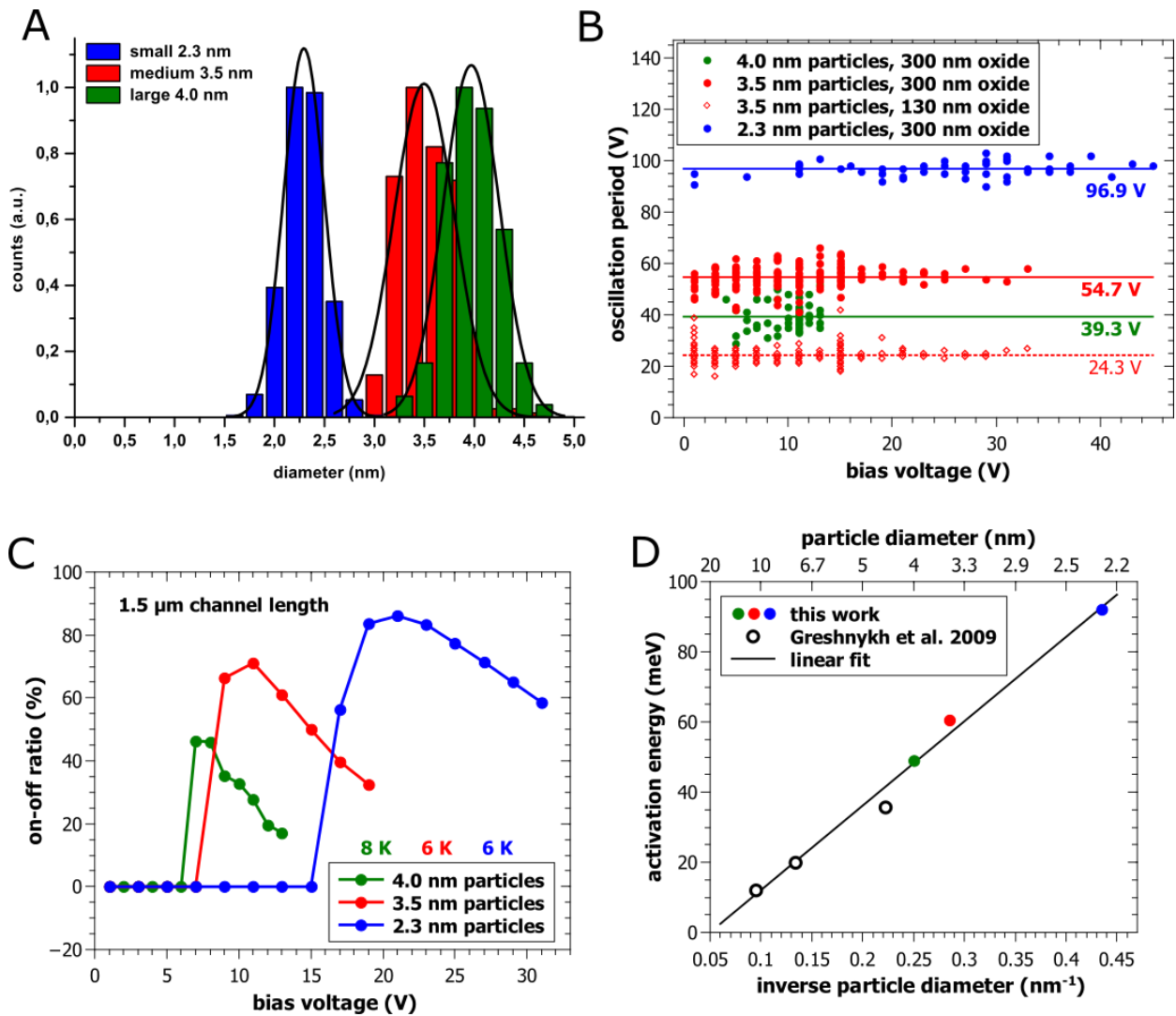

**Figure 6. Particle size comparison.** (**A**) Size distributions including Gaussian fits. (**B**) Period of oscillation for different channel lengths, temperatures and bias voltages. The horizontal lines mark the average value for each set of data points. The period seems to be only dependent on particle size and separation between channel and gate electrode. With these two parameters, it can be tuned over a wide voltage range. (**C**) Bias voltage dependent on/off ratios at low temperature for the three sizes. The best performance can be obtained by using small particles. (**D**) The activation energy is inversely proportional to the particle diameter. The results of this work supplement data from another publication.